\documentclass[conference]{IEEEtran}
\IEEEoverridecommandlockouts

\usepackage{times}
\usepackage{graphicx}
\usepackage{authblk}
\usepackage[cmex10]{amsmath}
\usepackage{amsthm}
\usepackage{amsmath}
\usepackage{algorithmic}
\usepackage{array}
\usepackage{bm,cite}
\usepackage{amssymb}

\usepackage[tight,footnotesize]{subfigure}
\usepackage{url,caption, color, enumerate, epsfig}

\title{Adversarial Machine Learning for Flooding Attacks on 5G Radio Access Network Slicing
\thanks{This effort is supported by the U.S. Army Research Office under contract W911NF-20-C-0055. The content of the information does not necessarily reflect the position or the policy of the U.S. Government, and no official endorsement should be inferred.}
}
	\author[1]{Yi Shi}
	\author[2]{Yalin E. Sagduyu}
		\affil[1]{\normalsize Virginia Tech, Blacksburg, VA 24061, USA}
		\affil[2]{\normalsize Intelligent Automation, Inc., Rockville, MD 20855, USA}

\begin{document}
\newcommand{\argmax}{\arg\!\max}
\maketitle

\begin{abstract}
Network slicing manages network resources as virtual resource blocks (RBs) for the 5G Radio Access Network (RAN). Each communication request comes with quality of experience (QoE) requirements such as throughput and latency/deadline, which can be met by assigning RBs, communication power, and processing power to the request. For a completed request, the achieved reward is measured by the weight (priority) of this request. Then, the reward is maximized over time by allocating resources, e.g., with reinforcement learning (RL). In this paper, we introduce a novel flooding attack on 5G network slicing, where an adversary generates fake network slicing requests to consume the 5G RAN resources that would be otherwise available to real requests. The adversary observes the spectrum and builds a surrogate model on the network slicing algorithm through RL that decides on how to craft fake requests to minimize the reward of real requests over time. We show that the portion of the reward achieved by real requests may be much less than the reward that would be achieved when there was no attack.  We also show that this flooding attack is more effective than other benchmark attacks such as random fake requests and fake requests with the minimum resource requirement (lowest QoE requirement). Fake requests may be detected due to their fixed weight. As an attack enhancement, we present schemes to randomize weights of fake requests and show that it is still possible to reduce the reward of real requests while maintaining the balance on weight distributions.
\end{abstract}

\section{Introduction}
5G promises unprecedented performance improvements in terms of rate, delay, and energy efficiency compared to 4G systems. An important design aspect towards this goal is resource allocation for network slicing that aims to multiplex and serve multiple virtualized and independent logical networks on the same physical network infrastructure of 5G \cite{KaloxylosSurvey, Foukas, TommasoRAN}. 5G radio access network (RAN) employs network slicing to manage its resources as virtual resource blocks (RBs), e.g., an RB may correspond to a frequency band. Then, the network resource allocation problem can be simplified by considering how to allocate virtual RBs without the need to focus on the physical resources to support RBs.
We consider a gNodeB that supports downlink communication requests from user equipments (UEs). A UE may generate  requests with different Quality of Experience (QoE) requirements for different types of network slices such as Enhanced Mobile Broadband (eMBB), Massive Machine Type Communications (mMTC), and Ultra Reliable Low Latency Communications (URLLC).
These QoE requirements are mapped to different requirements on RBs, as well as communication and processing powers at the gNodeB. Each request has its own priority (measured by a weight). If the gNodeB allocates resources to a request to meet its requirements, the reward is the weight of this request. If a request is not served, it is kept in a list of active requests until its deadline. This resource allocation approach aims to maximize the reward over time and can be used in near-real time RAN Intelligent Controller (Near-RT RIC)  to support micro-service-based applications called xApps.

Since future requests are unknown, an online algorithm is needed to make decisions based on current network status and requests. Machine learning can be effectively applied to solve complex wireless optimization problems by learning from spectrum data \cite{WirelessDL}. In particular, deep learning was studied for network slicing in \cite{NakaoML} for application and device specific identification and traffic classification, and in \cite{ThantharateML} for management of network load efficiency and network availability. As the training data may not be available, reinforcement learning (RL) was used for network slicing without requiring a prior model \cite{KooSlicingRL, LiSlicingRL, WangSlicingRL, Gursoy, CAMAD, Nassar20}. In our setting, the RL approach learns a model to predict the future reward (the weight of selected requests) for each state (available resources) and each action (admission of requests), and determines over time the optimal action to maximize the expected future reward.

With the success of applying machine learning (ML) to network slicing problems, there are also security concerns, specifically due to the attacks built upon adversarial ML. Adversarial ML studies the learning process in the presence of adversaries and expands the attack surface with new wireless attacks, e.g., exploratory (inference) attacks \cite{Terpek}, evasion (adversarial) attacks \cite{Larsson2, Gunduz, Kim1, Kim2,  Kim4}, causative (poisoning) attacks \cite{Sagduyu1}, membership inference attacks \cite{MIA}, Trojan attacks \cite{Davaslioglu1}, and signal spoofing attacks \cite{WiseMLSpoofing, YiSpoofing}. Recently, adversarial ML has been used to launch attacks on 5G such as attack on 5G spectrum sharing with incumbents, attack on 5G UE authentication \cite{5Gbc}, covert 5G communications \cite{Asilomar}, and attack on 5G network slicing (where the adversary aims to manipulate the underlying RL algorithm) \cite{NetSliceAttack}. In this paper, we consider a flooding type of resource starvation attack that has found applications in the networking domain, such as the TCP SYN flood attack. For network slicing, we formulate the flooding attack as the case that an adversary generates fake requests to consume network resources. This attack cannot be detected by simply monitoring the reward since the gNodeB can still achieve large reward as in the case of no attack. However, a great portion of this reward is associated with fake requests and the reward by real requests is much less than that without attack. Compared to conventional jamming (e.g., \cite{conventionaljamming}), the flooding attack is stealthier and more energy-efficient (especially for downlink traffic), since it only requires an adversary to send requests and ACK without a need of long transmissions.

The design challenge of flooding attack is how to generate fake requests for resources. If there is no limitation, the adversary can generate many requests to maximize its impact, but it can be easily detected and blocked. Thus, we assume an upper bound on the number of fake requests. The adversary needs to carefully design its requests for network resources and rewards such that the impact of these requests is maximized. For that purpose, we design an RL solution for the adversary that uses network resources as the state, the generated fake requests as the action, and the reward achieved by fake requests as the reward for RL. This solution aims to maximize the total reward for fake requests over time, which in turn minimizes the remaining reward for real requests. We show that this flooding attack reduces the reward for real requests significantly more than two benchmark attacks, namely random fake requests and fake requests with lowest QoE requirement (or minimum resource requirement). Although we mainly launch the flooding attack on an RL based network slicing scheme, we show that it can also target other network slicing schemes such as myopic, first come first served (FCFS), or random selection.

The best strategy (in terms of minimizing the reward of real requests) is to generate fake requests with large weights, but they may be easily detected by checking their weights over time. To overcome this defense, we design attack schemes with variable weights and show that we can set a close to uniform weight distribution on fake requests such that they cannot be easily detected based on their weights and the reward of real requests can still be reduced significantly.

The rest of the paper is organized as follows.
Section~\ref{sec:RLNS} describes the resource allocation schemes for network slicing.
Section~\ref{sec:attack} presents the flooding attack that aims to minimize the gNodeB's performance for real communication requests.
Section~\ref{sec:eval} evaluates the attack performance under different settings and designs.
Section~\ref{sec:conc} concludes this paper.

\section{Resource Allocation for Network Slicing}
\label{sec:RLNS}

\begin{figure}
	\centering
	\includegraphics[width=0.88\columnwidth]{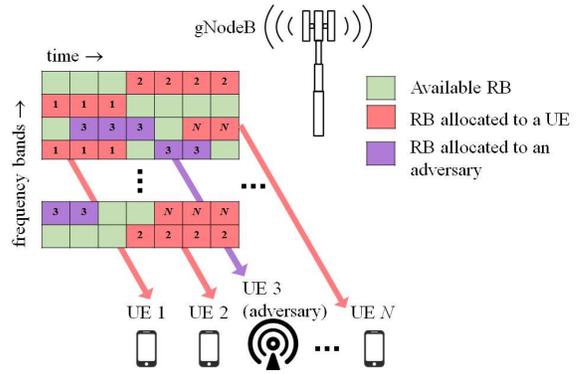}
	\caption{System model for the attack on 5G network slicing.}
	\label{fig:sys}
\end{figure}

Resource allocation for network slicing can be optimized by RL. We follow the RL approach in \cite{CAMAD} as an example.
Fig.~\ref{fig:sys} shows multiple 5G UEs sending requests over time with different rate, processing power, latency (deadline) and lifetime requirements and priority weights. The 5G gNodeB selectively serves some of these requests competing for resources. If a request is selected, resources are allocated to meet the requirements. Otherwise, it will be kept in a waiting list until its deadline expires. The 5G gNodeB selects requests to maximize the reward, i.e., the total weight of served requests over a time period.
An adversary launches the flooding attack to generate fake requests such that the total weight of served real requests over a time period can be minimized.
Denote the set of active requests (newly arrived requests and previously arrived requests that stay in the waiting list) at time $t$ as $A(t)$ for each time $t$. RBs, communication and processing powers are allocated to meet the requirements of admitted requests in $A(t)$. The rate and processing power requirements of UE $i$ for its request $j$ are given by
\begin{eqnarray}
D_{ij} \geq d_{ij} x_{ij} (t), \: \:
P_{ij} \geq p_{ij} x_{ij} (t), \: (i,j) \in A(t), \label{eq:DP}
\end{eqnarray}	
where $D_{ij}$ is the achieved data rate, $d_{ij}$ is the minimum required rate,
$P_{ij}$ is the assigned processing power, $p_{ij}$ is the minimum required processing power (measured by the percentage of CPU usage), and
$x_{ij} (t)$ is the binary indicator on whether UE $i$'s request $j$ is satisfied at time $t$.
At any time, the total $P_{ij}$ of selected requests is no more than $1$. $D_{ij}$ measured in bps depends on the assigned bandwidth $F_{ij}$ and the modulation coding scheme used for communications from gNodeB to UE $i$, and is approximated as $D_{ij} = c \;\cdotp K_{ij} \cdotp (1-\textit{BER}_{ij})$ \cite{5GNRStd1},
where $K_{ij}$ is the number of allocated RBs and $\textit{BER}_{ij}$ is the bit error rate of UE $i$ for its request $j$,
and constant $c$ is approximately $12.59 \cdot 10^6$ when a single-antenna UE uses QPSK modulation, $60$ kHz subcarrier spacing and $10$ MHz bandwidth.
The constraints of RB assignments are
\begin{eqnarray}
\sum_{i,j} F_{ij} x_{ij}(t) \leq  F(t), (i,j) \in A(t), \label{eq:assign-f} 		
\end{eqnarray}
where $F(t)$ represents the available RBs of the gNodeB at time $t$ (resources that are assigned previously to some requests and not released yet become temporarily unavailable). By considering the optimization problem for a time horizon, the resources are updated from time $t-1$ to time $t$ as
\begin{eqnarray}
F(t)  =  F(t-1)+F_r (t-1)-F_a (t-1), \label{eq:update-f}
\end{eqnarray}			
where $F_r (t-1)$ and $F_a (t-1)$ are released and allocated resources on frequency at time $t-1$. Each request has a lifetime $l_{ij}$ and if it is selected at time $t_s$ (namely, the service starts at time $t_s$), this request will end at time $t_s+l_{ij}$. The released and allocated resources at time $t$ are given by
\begin{eqnarray}
F_r (t) =  \sum_{(i,j) \in R(t)} F_{ij}, \:\:\:\:
F_a (t) =  \sum_{i,j} F_{ij} x_{ij}(t), \label{eq:allocate-f}	\end{eqnarray}			
respectively, where $R(t)$ denotes the set of requests ending at time $t$.
Then, the optimization problem is given by
\begin{eqnarray}
\max_{x_{ij}(t)} \sum_t \sum_{(i,j)}  w_{ij} x_{ij}(t), \:\: (i,j) \in A(t)
\label{eq:opt2}
\end{eqnarray}
subject to (\ref{eq:DP})--(\ref{eq:allocate-f}), where $w_{ij}$ is the weight for UE $i$'s request $j$ to reflect its priority. Without knowing future requests, the model-free RL algorithm solves (\ref{eq:opt2}) by making decisions using an online learned policy that determines an action for current state for the gNodeB. In this paper, we consider Q-learning from \cite{CAMAD}.
The reward at time $t$ is $w_{ij}$ if UE $i$'s request $j$ is satisfied, i.e., $x_{ij}(t) = 1$. Actions assign resources to each request at time $t$. Multiple actions can be taken at the same time instance if there are sufficient resources.
The states at $t$ are the remaining RBs and communication and processing powers. Given by (\ref{eq:update-f})-(\ref{eq:allocate-f}), state transition at time $t$ is driven by allocating resources for requests granted at time $t$ and releasing resources after lifetimes of some active services expire at time $t$. Note that the flooding attack can be also launched on other network slicing schemes. We consider the following schemes for comparison purposes. The myopic scheme aims to maximize the reward for the current time (without considering future rewards). The FCFS scheme aims to allocate resources based on the arrival time of requests. The random scheme allocates resources to some randomly selected requests.

\section{Reinforcement Learning based Flooding Attack for Network Slicing}
\label{sec:attack}
An adversary attacks the 5G RAN network slicing by generating fake requests for network slices. If these fake requests are selected and network resources are allocated to them, fewer resources will be left for real requests from legitimate users.
As a consequence, although a gNodeB may still achieve a high reward for many granted requests, the actual reward corresponding to real requests may be a smaller option of it. We consider a practical constraint that the adversary has a limited rate of generating fake requests to avoid being detected. In particular, we can set the same rate of request generation for the adversary and legitimate users. Thus, it is important for the adversary to generate fake requests with two objectives:
\begin{itemize}
\item Objective 1: fake requests are satisfied (namely, resources are allocated to them) with high probability.
\item Objective 2: fake requests occupy resources so that real requests cannot be satisfied due to no resources.
\end{itemize}
For the first objective, a fake request should ask for a smaller portion of resources (to avoid being rejected due to insufficient resources) and have the maximum reward (to have high priority). For the second objective, a fake request should consume the majority of the resources such that the remaining resources are not sufficient for real requests. Thus, the ideal setting on weights in fake requests is the largest weight while the required resource should not be too large (to ensure that a fake request can be satisfied) or too small (to ensure that a significant portion of the total resources can be occupied by fake requests).
Therefore, the key step in the flooding attack is determining resources to be specified by fake requests.

The resources include the number of available RBs, the remaining processing power (memory), and the remaining communication (transmit) power. The adversary may sense the spectrum and aim to detect available RBs. However, it cannot know the remaining processing and communication powers. There is no need to consume all of these resources to prevent serving real requests, since a request cannot be satisfied if any of its resource requirements is not met.
Thus, the adversary can request minimum processing and communication powers such that its requests will not be rejected due to the lack of these resources. On the other hand, the adversary needs to determine RB requirements in its requests. To make decisions, we design a Q-learning algorithm for the adversary as follows.
\begin{itemize}
\item The state is the number of available RBs.

The action is to select how many RBs to be assigned in a fake request ($0$ means no request is made). The number of potential actions is $n_a +1$, where $n_a$ is the number of available RBs.

\item The reward is the number of served fake requests (or the total reward of served fake requests).
\end{itemize}
The Q-table maps (state, action) to reward. To initialize this table, if a fake request is generated, the corresponding entry is set as $1$, otherwise the entry is set as $0$.
The adversary applies Q-learning to update this table and take actions based on this table. The adversary generates a fake request only if the rate of fake requests so far is below the expected rate, which can be set the same as the rate of real requests from other (legitimate) users. Under the flooding attack, we measure both the total reward (including the reward for both real and fake requests) and the real reward (including the reward for real requests only).

\section{Performance Evaluation}
\label{sec:eval}

\subsection{Flooding Attack Results}

Suppose that the gNodeB receives requests from three UEs. For each UE, requests arrive with rate of $0.5$ per slot. The adversary also generates fake requests at this rate. Here, a slot corresponds to each time block which is $0.23$ ms long with $60$ kHz subcarrier spacing. For each request, weight, lifetime, and deadline are randomly assigned in $[1,5]$, $[1,10]$ slots, and $[1,20]$ slots, respectively. The signal-noise-ratio (SNR) is selected randomly from $[1.5,3]$. The total frequency is $10$ MHz and is split into $11$ bands, i.e., there are $11$ RBs. In addition to the Q-learning-based attack, we also consider the case of no attack and two benchmark attacks:
\begin{itemize}
	\item \textit{Random attack}: The adversary generates fake requests with random requirements on RBs.

	\item \textit{Minimum resource (MinRes) attack}: The adversary generates fake requests with the lowest QoE requirement, which in turn requires the minimum number of RBs, i.e., always one RB is required.
\end{itemize}

\begin{table}
	\caption{Performance comparison of flooding attack using Q-learning and other attack schemes.}
	\centering
	{\small
		\begin{tabular}{c|c|c}
Algorithm & Total reward & Real reward \\ \hline \hline
Q-learning & 2593 & 523 \\ \hline
MinRes & 2769 & 614 \\ \hline
Random & 1905 & 1630 \\ \hline
No attack & 1783 & 1783
		\end{tabular}
	}
	\label{table:1}
\end{table}

We assume that the adversary launches its attack (Q-learning, MinRes, or Random) over $10000$ slots. The benchmark of no attack case is also run over $10000$ slots. The achieved reward is measured for the last $1000$ slots. For network slicing, we first consider the RL based scheme.
Table~\ref{table:1} shows the performance of different attacks and the case of no attack. If there is no attack, i.e., all requests are real, the total reward and the real portion of it are the same. All attacks increase the total reward, since there are some additional fake requests with high reward, but the real portion of the reward is less than the total (all real) reward achieved under no attack.
The random attack does not work well and only slightly reduces the real reward (from $1783$ to $1630$). The MinRes attack always generates fake requests with the minimum required resource. Although this increases the probability of being selected by the gNodeB for service, the occupied resource is also minimized. Hence, it significantly reduces the real reward to $614$ but it is not as effective as the Q-learning attack, which reduces the real reward to $523$.

We also measure the total reward of all fake requests generated under the Q-learning attack, which is $2440$. The total reward of served fake requests is $2593-523=2070$, i.e., most of fake requests are served and occupy some resources. Thus, the Q-learning attack is efficient in terms of the ratio between served and generated requests. The total reward asked for real requests is $4594$, while the total reward of served real requests is only $523$ under the Q-learning attack, i.e., only a small portion of real requests are served under the flooding attack, showing that the flooding attack is highly successful.

\begin{table}
	\caption{Performance comparison of different network slicing schemes under the flooding attack.}
	\centering
	{\small
		\begin{tabular}{c|c|c|c}
Network slicing & No attack & \multicolumn{2}{c}{Flooding attack} \\ \cline{3-4}
scheme & & Total reward & Real reward \\ \hline \hline
Q-learning & 1786 & 2593 & 523 \\ \hline
Myopic & 1422 & 2408 & 653 \\ \hline
FCFS & 1416 & 2369 & 429 \\ \hline
random & 1318 & 2363 & 483
		\end{tabular}
	}
	\label{table:ns}
\end{table}

This flooding attack can be launched against other network slicing schemes including the myopic, FCFS, and random  schemes described in Section~\ref{sec:RLNS}. Table~\ref{table:ns} shows that the (reward) performance of all these schemes drop significantly under the flooding attack. In particular, the FCFS and random schemes have worse performance than the Q-learning based scheme regardless there is a flooding attack, or not. One interesting result is that the myopic scheme has worse performance compared to the Q-learning based scheme when there is no attack, but under the flooding attack, the myopic scheme achieves better reward, namely $653$, compared to the reward $523$ achieved by Q-learning based scheme. The reason is that the Q-learning based scheme aims to maximize the expected total reward, including both current reward and future reward. When there are fake requests with high rewards, the Q-learning algorithm tends not to allocate resources to real requests if their reward is not high. On the other hand, there is no such issue in the myopic scheme since it only considers the current reward.

\subsection{Impact of System Parameters}

Now we check the effect of system parameters on the flooding attack performance. Table~\ref{table:rate} shows the results when we vary the rate of fake requests, $r_f$. The total reward increases first with $r_f$ due to high reward of fake requests, while the real reward decreases due to the increasing portion of fake requests. If $r_f \geq 0.7$ request per slot, the total or real reward does not change, i.e., $r_f = 0.7$ is sufficient for the best attack.

\begin{table}
	\caption{The effect of the generation rate of fake requests, $r_f$.}
	\centering
	{\small
		\begin{tabular}{c|c|c}
$r_f$ & Total reward & Real reward \\ \hline \hline
0 & 1783 & 1783 \\ \hline
0.1 & 2077 & 1587 \\ \hline
0.2 & 2327 & 1352 \\ \hline
0.3 & 2512 & 1032 \\ \hline
0.4 & 2605 & 700 \\ \hline
0.5 & 2593 & 523 \\ \hline
0.6 & 2620 & 440 \\ \hline
$\ge$ 0.7 & 2612 & 372
		\end{tabular}
	}
	\label{table:rate}
\end{table}

Table~\ref{table:rb} shows the results when we vary the number of RBs, $n_r$. The reward when there is no attack increases first with $n_r$ due to more RBs, and then changes within certain range due to randomness in limited number of user requests. The total reward under attack shows the same trend.  The real reward under attack first decreases with $n_r$ and then stays within certain range. To understand the trend better, we assess the ratio between real reward under attack and the reward when there is no attack. By increasing $n_r$, this ratio first decreases as a fake request can attack more RBs, and then stays within certain range since (i) the number of fake requests is limited and (ii) there is some randomness in generated fake requests.

\begin{table}
	\caption{The effect of the number of RBs, $n_r$.}
	\centering
	{\small
		\begin{tabular}{c|c|c|c|c}
$n_r$ & No attack & \multicolumn{2}{c|}{Flooding attack} & Ratio (\%) \\ \cline{3-4}
 &  & Total reward & Real reward & \\ \hline \hline
5 & 1526 & 2133 & 858 & 56.23 \\ \hline
6 & 1605 & 2285 & 805 & 50.16 \\ \hline
7 & 1786 & 2477 & 737 & 41.27 \\ \hline
8 & 1730 & 2596 & 646 & 37.34 \\ \hline
9 & 1870 & 2619 & 609 & 32.57 \\ \hline
10 & 1902 & 2644 & 604 & 31.76 \\ \hline
11 & 1783 & 2593 & 523 & 29.33 \\ \hline
12 & 2018 & 2723 & 528 & 26.16 \\ \hline
13 & 1964 & 2677 & 577 & 29.38 \\ \hline
14 & 1987 & 2661 & 506 & 25.47 \\ \hline
15 & 1899 & 2708 & 573 & 30.17
		\end{tabular}
	}
	\label{table:rb}
\end{table}

Table~\ref{table:user} shows the results when we vary the number of users, $n_u$. By increasing $n_u$, the reward when there is no attack increases due to more requests to be selected for service. The total reward and the real reward under attack show the same trend. When $n_u$ is large, the total reward under attack is less than the reward when there is no attack. The reason is that the adversary generates fake requests with non-minimum RBs while with many users it is likely that there are requests with the same reward and minimum RBs. Then, the total reward by selecting some fake requests may be less than the case of not selecting fake requests. The ratio of real reward over reward under no attack increases with $n_u$ due to more real requests competing with fake requests.

\begin{table}
	\caption{The effect of the number of users, $n_u$.}
	\centering
	{\small
		\begin{tabular}{c|c|c|c|c}
User & No attack & \multicolumn{2}{c|}{Flooding attack} & Ratio (\%) \\ \cline{3-4}
 &  & Total reward & Real reward & \\ \hline \hline
3 & 1783 & 2593 & 523 & 29.33 \\ \hline
4 & 2239 & 2717 & 652 & 29.12 \\ \hline
5 & 2306 & 2749 & 709 & 30.75 \\ \hline
6 & 2681 & 2856 & 955 & 35.62 \\ \hline
7 & 2592 & 2850 & 890 & 34.34 \\ \hline
8 & 2865 & 2854 & 954 & 33.30 \\ \hline
9 & 2896 & 2859 & 1029 & 35.53 \\ \hline
10 & 3108 & 2918 & 1108 & 35.65 \\ \hline
20 & 3540 & 3065 & 1480 & 41.81 \\ \hline
50 & 3859 & 3360 & 2100 & 54.42
		\end{tabular}
	}
	\label{table:user}
\end{table}

Table~\ref{table:snr} shows the results when we vary the SNR for users. By increasing the SNR,
all rewards (real or total) increase with and without flooding attack. In particular, the ratio of real reward over the reward under no attack increases due to better channels available for users. In this case, it is easier to serve a real request by meeting rate requirements, i.e., it is more challenging for the attack to deny service to a real request by consuming resources.

\begin{table}
	\caption{The effect of the SNR for users.}
	\centering
	{\small
		\begin{tabular}{c|c|c|c|c}
SNR & No attack & \multicolumn{2}{c|}{Flooding attack} & Ratio (\%) \\ \cline{3-4}
 &  & Total reward & Real reward & \\ \hline \hline
low & 1557 & 2436 & 286 & 18.37 \\ \hline
medium & 1783 & 2593 & 523 & 29.33 \\ \hline
high & 1900 & 2695 & 625 & 32.89
		\end{tabular}
	}
	\label{table:snr}
\end{table}

Finally, we evaluate the effect of the adversary's observation error on available RBs, in terms of false alarm (available RBs are detected as unavailable) and misdetection (unavailable RBs are detected as available). We find that this effect is not significant. Even for significant errors up to $20$\%, the change on real reward is at most $7.07$\%. Hence, the flooding attack is not very sensitive to errors in spectrum sensing.

\subsection{Attack Extensions with Enhanced Weight Distribution}

The flooding attack that we consider so far generates fake requests with the largest weight (LW) to maximize the probability that they are selected by the gNodeB. A defense scheme may detect the largest weight in requests from an adversary and then discard these requests. Against such a defense, we extend the flooding attack with the following schemes to increase randomness of weights.
\begin{itemize}
\item \textit{Uniform weight (UW)}:  Weights in fake requests are uniformly randomly assigned.

\item \textit{Uniform large reward (ULW)}: Weights in fake requests are uniformly randomly assigned as large values $4$ or $5$.

\item \textit{Resource dependent weight (RDW)}: The flooding attack requests should be accepted when remaining resources are large, since it is more likely that real requests can be accepted if no attack. Thus, the adversary can generate more fake requests with large weights if remaining resources are large. Otherwise, the adversary can generate more fake requests with small weights. The resource dependent weight distributions are selected such that the overall weight distribution remains uniform, i.e., $\frac{1}{F} \sum_i p_{ij} = \frac{1}{W}$, where $F$ is the number of total available RBs, $p_{ij}$ is the probability of weight is $j$ when the number of remaining RBs is $i$, and $W$ is the number of different weights. One such distribution is shown in Table~\ref{table:biased}.

\begin{table}
	\caption{Distribution of weights in network slicing requests.}
	\centering
	{\small
		\begin{tabular}{c|c|c|c|c|c|c|c}
$j$ $\backslash$ $i$ & 1,2 & 3,4 & 5 & 6 & 7 & 8,9 & 10,11 \\ \hline \hline
1 & 0.5 & 0.4 & 0.2 & 0.2 & 0 & 0 & 0 \\ \hline
2 & 0.4 & 0.4 & 0.3 & 0.2 & 0.1 & 0 & 0 \\ \hline
3 & 0.1 & 0.2 & 0.4 & 0.2 & 0.4 & 0.2 & 0.1 \\ \hline
4 & 0 & 0 & 0.1 & 0.2 & 0.3 & 0.4 & 0.4 \\ \hline
5 & 0 & 0 & 0 & 0.2 & 0.2 & 0.4 & 0.5
		\end{tabular}
	}
	\label{table:biased}
\end{table}

\item \textit{Resource dependent large weight (RDLW)}: We limit the weights in fake requests to large values $4$ or $5$, and choose weight distributions such that $\frac{1}{F} \sum_i p_{ij} = 0.5$ for $j=4$ or $5$. One such distribution is shown in Table~\ref{table:bhr}.

\begin{table}
	\caption{Distribution for high weight.}
	\centering
	{\small
		\begin{tabular}{c|c|c|c|c|c|c|c}
$j$ $\backslash$ $i$ & 1 & 2,3 & 4,5 & 6 & 7,8 & 9,10 & 11 \\ \hline \hline
1 & 0 & 0 & 0 & 0 & 0 & 0 & 0 \\ \hline
2 & 0 & 0 & 0 & 0 & 0 & 0 & 0 \\ \hline
3 & 0 & 0 & 0 & 0 & 0 & 0 & 0 \\ \hline
4 & 1 & 0.9 & 0.8 & 0.5 & 0.2 & 0.1 & 0 \\ \hline
5 & 0 & 0.1 & 0.2 & 0.5 & 0.8 & 0.9 & 1
		\end{tabular}
	}
	\label{table:bhr}
\end{table}

\item \textit{Adjusted weight (AW)}: The adversary adjusts weights dynamically based on whether a fake request is selected or not. If a fake request is selected, the adversary wants to have another fake request being selected to occupy resources, otherwise there is no such need. Thus, the weight in requests should be increased (bounded by the largest reward) if a fake request is selected, otherwise the weight should be decreased (bounded by the smallest reward). We consider three adjustment approaches.
    \begin{itemize}
     \item AW 1: increase the weight in fake requests by $1$ if a fake request is selected or decrease it by $1$ otherwise,
     \item AW 2: increase the weight in fake requests to the largest value if a fake request is selected or decrease it by $1$ otherwise, or
     \item AW 3: increase the weight in fake requests to the largest value if a fake request is selected or decrease it by $1$ with a probability otherwise.
     \end{itemize}
\end{itemize}

Table~\ref{table:2} shows results for these attack extensions. First of all, since the adversary aims to sustain uniform distribution among different weights in its requests, the probability of selecting a fake request is reduced and thus the real reward under all these attacks is larger than the LW attack case where we set the weight in fake requests as the largest value. Under the flooding attack with UW, the real reward is $1161$ while the real reward under the flooding attack with ULW is $828$.
The real reward achieved under the flooding attack with RDW is $1008$, which is still much higher than the real reward $523$ when we fix the reward as its largest value.
The real reward under the flooding attack with RDLW is $677$, which is close to $523$ when we fix reward as $5$.
For the flooding attack with AW, AW 1 yields a real reward $1417$, which is high, as it turns out that most weights for fake requests are selected as $1$. When AW 2 is used, the real reward becomes $1008$ and we find that the distribution of weights is almost a uniform distribution. When we use AW 3 with the probability set as $0.4$, which also yields an approximately uniform distribution for reward in fake requests, the achieved real reward is $1146$. Overall, the flooding attack can reduce the real reward significantly while keeping close to uniform weight distribution such that it is difficult to detect fake requests by checking their weights.

\begin{table}
	\caption{Attack extensions with different weights in fake requests.}
	\centering
	{\small
		\begin{tabular}{c|c|c}
Algorithm & Total reward & Real reward \\ \hline \hline
LW & 2593 & 523 \\ \hline
UW & 2328 & 1161 \\ \hline
ULW & 2621 & 828 \\ \hline
RDW & 2597 & 1008 \\ \hline
RDLW & 2689 & 677 \\ \hline
AW 1 & 1798 & 1417 \\ \hline
AW 2 & 2389 & 1008 \\ \hline
AW 3 & 2245 & 1146
		\end{tabular}
	}
	\label{table:2}
\end{table}

\section{Conclusion}
\label{sec:conc}
We presented a flooding attack on 5G RAN slicing, where an adversary generates fake network slicing requests to consume available resources and minimize the reward for real requests. We developed an RL based attack to generate fake requests and showed that it is more effective than generating fake requests randomly or with minimum resource requirement. Although we focused on attacking an RL based network slicing, we showed that the flooding attack is effective against other network slicing schemes. We designed weight distribution schemes for fake requests such that they cannot be detected by their weights, and showed that the flooding attack using a close to uniform weight distribution is still effective. Our results indicate that 5G RAN slicing is highly vulnerable to flooding attacks that can significantly reduce the reward of real requests by starving resources with fake requests.

\end{document}